# Ionization of Cucurbiturils as a Pathway to More Stable Host-Guest Complexes

Vitaly V. Chaban, Eudes Eterno Fileti and Thaciana Malaspina

Instituto de Ciência e Tecnologia, Universidade Federal de São Paulo, 12231-280, São José dos Campos, SP, Brazil.

**Abstract**. Cucurbiturils are particularly interesting to chemists, because these macrocyclic molecules are suitable hosts for an array of neutral and cationic species. It is believed that the host-guest binding originated from hydrophobic interactions and ion-dipole interactions in the case of cationic guests. The fact that an elementary unit of cucurbiturils (glycoluril unit) consists of two fused imidazole rings, which ionize readily, has remained largely unnoticed up to now. This work reports ionized cucurbiturils and their binding to $C_{60}$ fullerene using versatile electronic-precision description. The methodology is based on density functional theory. We assert that cationization of cucurbiturils fosters $C_{60}$–cucurbituril due to polarization of electron density in $C_{60}$. Therefore, more stable host-guest complexes can be derived.

**TOC Graphic**

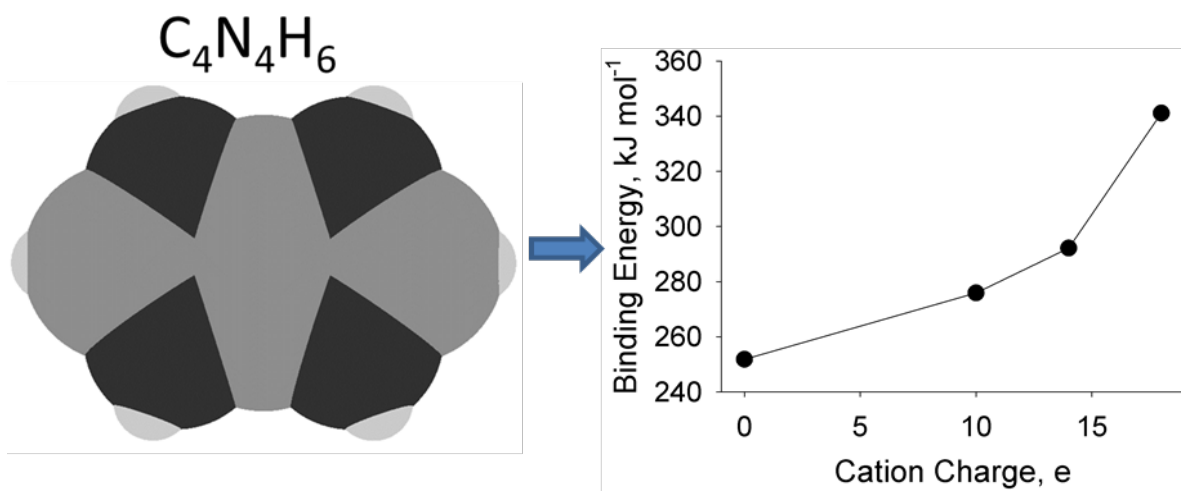

**Introduction**

The difference between a molecule and an ion (cation or anion) is just one or two electrons. Nevertheless, the ionization process significantly alters intermolecular interactions making them more electrostatic in nature. Anionization is normally an exothermic process with the first electron affinity ranging from units to hundreds of kilojoules per mole. Contrariwise, cationization requires external energy, 500-1500 kJ mol$^{-1}$, to be supplied. Simpler formation of anions, as compared to cations, implicitly stands beyond the fact that anions are commonly polyatomic, whereas most cations are monoatomic. The exception is a relatively versatile group of organic cations, which are based on the excessively protonated heterocycles.

Cucurbiturils (CBs) are well-known in supramolecular chemistry as molecular containers, which form stable complexes with a variety of guests, such as hydrocarbons, saccharides, drugs, amino acids, peptides and proteins.[1,2] These complexes are interesting in the context of nanoscale applications including controlled drug delivery,[3-5] molecular machines,[4-8] supramolecular polymers[9-11] and biomimetic systems.[3,12] CBs[n], where n is a number of units ranging between five and ten, represent a family of molecular containers. These compounds possess a hydrophobic internal cavity and a hydrophilic exterior. The oxygen atoms are located at the edge of the structure. Due to specific intra-molecular interactions, the oxygen atoms appear tilted inwards. Thus, a partially closed cavity, which is unique in many aspects, is established. Due to significant cavity size, even the smallest CBs can be employed as hosts for a variety of species. Neutral[7] and cationic entities[13] were previously observed inside the hydrophobic cucurbituril cavity.

The varying cavity size in CBs leads to well adjustable recognition properties. That is, different CBs accomodate different molecules and ions. CBs are usually functionalized with anions, such as sulfonate and carboxylate moieties, which change the key sites of supramolecular interaction. Inclusion of cationic guests and anionic functionalization of CBs aim towards solubility increase in common solvents.[14]

Because of good biocompatibility and strong affinities toward some guest molecules, CBs have also been utilized to host a variety of biomedical applications as drug delivery, biological imaging and photodynamic therapy.[15-22] Collins and co-workers used CB[7] to encapsulate anticancer drug dinuclear platinum complex and studied its reaction rate, cytotoxicity, and interaction with DNA.[15-17] They also used CB[6], CB[7], and CB[8] to complex albendazole and increased its aqueous solubility by 2000-fold.[18] Host-guest complexation between CBs and fluorescence dyes has been also used to overcome drawbacks in biological imaging. For instance, the host-guest interaction between CB[7] and Hoechst dyes for fluorescence emission enhancement and responsiveness to external pH changes was investigated.[19, 20] Methylene blue (MB), a photosensitizer, was encapsulated into the cavity of CB[7], forming a host-guest complex with high binding constant.[21] The aggregation of MB was avoided by such encapsulation even at high concentration, and CB[7] could also protect MB from oxygen attack, reserving its photodynamic activity. That work still showed the significance of using the CB host-guest complex for reserving photosensitizer for photodynamic therapy.[21] The photodynamic properties of a supramolecular cucurbit[8]uril–fullerene complex have been studied by Jiang and Li, who attributed HeLa cell death mainly to the damage of membrane phospholipids and proteins.[22]

Recently, a new class of host-guest structures involving cucurbiturils and fullerenes was developed.[23] Fileti and coworkers investigated thermodynamics stability and structure of the $C_{60}$ fullerene and CB[9]. Since an internal diameter of the CB[9] is similar to the van der Waals diameter of the $C_{60}$ fullerene, an unusually stable inclusion complex was obtained and characterized.

The host-guest properties of the conventional molecular cucurbiturils are being actively studied with neutral and cationic guests.[7, 9, 10, 13] It remained unnoticed up to now that an elementary unit of all cucurbiturils consists of two fused imidazole rings. These rings can be transformed into cations using conventional techniques. In this work, we report cationization of CB[5] and CB[9] and investigate binding of the resulting species with the $C_{60}$ fullerene.

**Methodology**

The results reported in this work are based on multiple electronic structure calculations of neutral and charged CB[5] and CB[9] and four complexes (q=0, q=+10e, q=+14e, q=+18e) of CB[9] with the $C_{60}$ fullerene. The solvent effects are omitted in the present work, thus optimized geometries obey an isolated molecule (ion) approximation.

Hybrid density functional theory (HDFT) calculations were used to derive electronic density distributions in the neutral and charged [CB[9]+$C_{60}$] host-guest complexes. The exchange energy was combined with the exact energy from the Hartree-Fock theory. We used the M06 HDFT functional[24] as a well-established and reliable one. Coupled cluster calculations using single and double substitutions from the Hartree-Fock determinant and including triple excitations non-iteratively[25] were employed to compute highly accurate thermodynamics potentials. The 6-311G Pople-type basis set where polarization and diffuse functions were supplemented to all atoms, except hydrogen atoms, was applied. No pseudopotentials were applied. That is, all electrons were considered explicitly at all calculation steps. The Grimme's empirical dispersion correction was applied to adjust resulting energies and forces.[26] This correction constitutes an important supplement, because the complexes involving cucurbiturils are stable largely thanks to hydrophobic interactions.

The geometry of each many-body electron-nucleus system was optimized by the Berny geometry optimization algorithm. The wave function convergence criterion at every self-consistent field (SCF) step was set to $10^{-6}$ Hartree. No additional techniques to enhance the convergence were applied. Most SCF cycles involved 10 or less iterations that is acceptable from the computational efficiency viewpoint. Hirshfeld partitioning scheme[27, 28] was used to derive point charges to be used in the subsequent analysis. In this scheme, pro-atomic densities are derived from computations on the neutral atoms by simply averaging the atomic electron density over the angular degrees of freedom.

The implementation of the outlined electronic structure methods in GAMESS[29] was employed.

**Results and Discussion**

One can easily notice that the glycoluril unit (an elementary unit of all cucurbiturils) represents basically two fused imidazole rings. The only difference is that one carbon atom chemically binds an oxygen atom, instead of a hydrogen atom. The C=O group can be reduced using conventional methods of synthetic organic chemistry. Similarly to imidazole, the resulting particle, $C_4N_4H_6$ (Figure 1), must give two electrons away (one from each ring) readily, therefore producing the doubly-charged cation, $[C_4N_4H_6]^{2+}$.

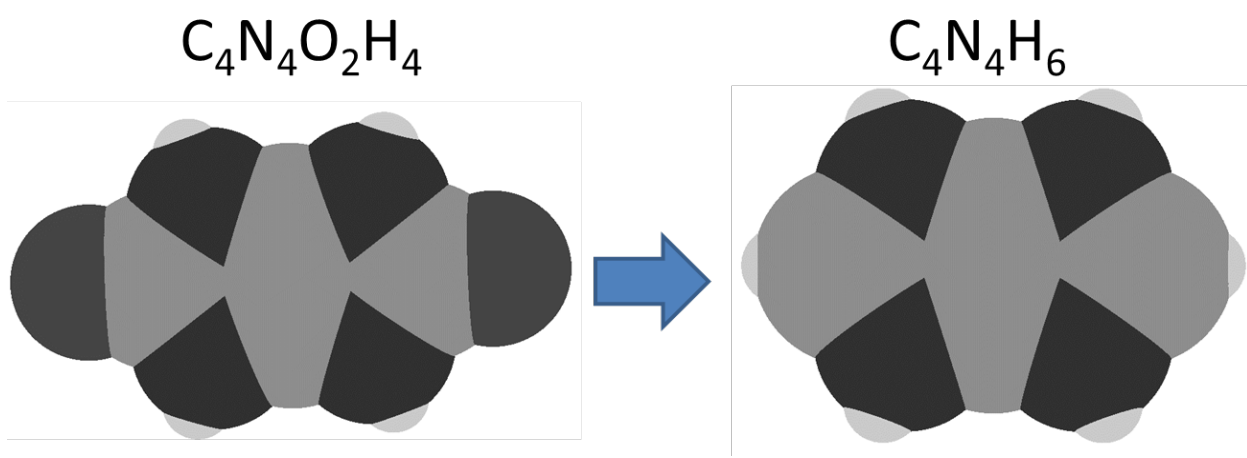

**Figure 1**. Scheme of reduction of the glycoluril unit. The resulting unit exhibits +2e positive charge and singlet multiplicity. The depicted geometries were optimized using the Berny algorithm. Hydrogen atoms are white, oxygen atoms are black. The size of each atom is approached to its experimental van der Waals radius.

To obtain a numerical evidence about stability of such doubly-charged cations, we computed thermodynamics potentials – enthalpy, entropy, and Gibbs free energy – for the corresponding chemical reaction at 300 K and 1 bar (Table 1). Note that these reaction schemes are simplified, whereas multiple synthetic approaches to perform the carbonyl group reduction are known. We consider both a simplified fragment $C_4H_4N_4O_2$ and $C_4N_4(CH_3)O_2$, where the methyl groups passivate nitrogen atoms, instead of hydrogen atoms. The imidazolium cations are stabilized by the two alkyl groups, which are linked to the two nitrogen atoms of the imidazole ring. Recall that nitrogen atoms in cucurbiturils are connected via methylene groups. Therefore, consideration of methyl groups in the glycoluril unit, instead of just hydrogen atoms, is fully rational. Consideration of thermodynamics potentials for the constructed chemical transformations (Table 1) is interesting in many aspects. It can suggest the most energetically favorable pathway for the CB[n] ionization. Note, however, that most energetically favorable pathway is not always easiest in implementation, since we do not consider all possible reaction stages and all possible energetic barriers.

Both reactions are energetically permitted. The free energy change is in line with that for other redox reactions. Recall that reduction of each C=O site is a separate process, therefore the listed energies may be divided by two for clearer comparison. The alteration of entropy at 300 K and 1 bar is negative, but gain of enthalpy compensates this partial energy decay. Compound $C_4N_4(CH_3)_4O_2$ reduces more easily than compound $C_4N_4H_4O_2$, although the relative difference is nearly marginal. We assume that the methyl groups play an important role in decreasing chemical reactivity and hence stability of the cation. As anticipated, the resulting doubly-charged cations constitute stable chemical entities.

**Table 1**. Thermodynamics potentials for ionization of the glycoluril derivatives at 300 K

| Compounds | | Thermodynamics Potentials | | |
|---|---|---|---|---|
| Reactants | Products | $\Delta H$, kJ mol$^{-1}$ | $\Delta S$, J mol$^{-1}$ K$^{-1}$ | $\Delta G$, kJ mol$^{-1}$ |

| | | | | | | |
|---|---|---|---|---|---|---|
| $C_4H_4N_4O_2$ | $H^+/H_2$ | $[C_4H_6N_4]^{2+}$ | $H_2O$ | -1456 | -149 | -1411 |
| $C_4N_4(CH_3)O_2$ | $H^+/H_2$ | $[C_4N_4(CH_3)H_2]^{2+}$ | $H_2O$ | -1487 | -143 | -1444 |

Stability of the the cations was additionally corroborated in the case of cucurbiturils. We selected the five-unit ring, CB[5] for this study. Note that here we consider only neutral and ionized CB[5] without any guest molecule. Figure 2 summarizes ionization potentials and electron affinities computed for CB[5], where all oxygen atoms were substituted by hydrogen atoms. The resulting charge of the cation equals to +10e, since each unit can maintain +2e. Electron affinities are always smaller than ionization energies, since acceptance of an additional electron is more energetically favorable than an electron loss. Energy of both types depicted in Figure 2 is sufficiently high to prohibit spontaneous transformation of the $CB[5]^{10+}$ cation. The electrostatic charge of +10e supports the most optimal structure of the given many-body electron-nucleus system.

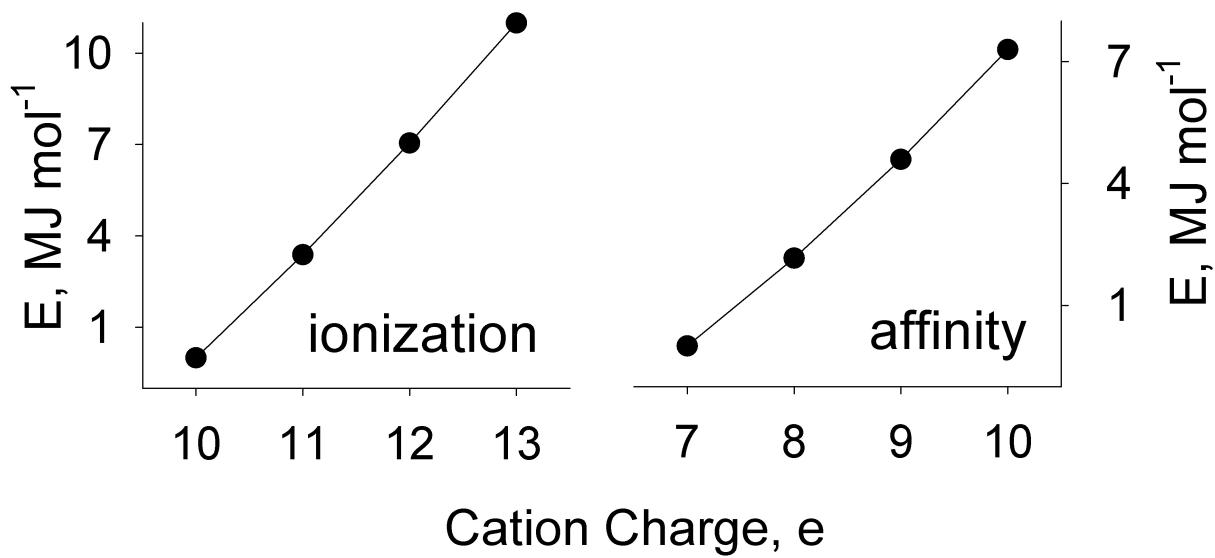

**Figure 2**. Ionization energies (left) and electron affinities (right) of the $CB[5]^{10+}$ cation. High positive energies (note that the energies are given in megajoules per mole) in both properties indicate stability of the developed ion. We used a smaller cucurbituril for this study (5 units), because fullerene (guest molecule) is absent in this system. Therefore, the host cavity can be smaller for a more computationally efficient investigation.

The cavity of cucurbit[9]uril and spherical (Figure 3). Moreover, it exhibits an exactly suitable size to encapsulate the $C_{60}$ fullerene. The cationization of CB[9] does not modify these features. Note that the geometries of the depicted host-guest complexes were thoroughly optimized following the potential energy surface. The stability of the complex suggests that the potential energy surface was not altered drastically as a result of significant (+10e, +14e, +18e) ionization.

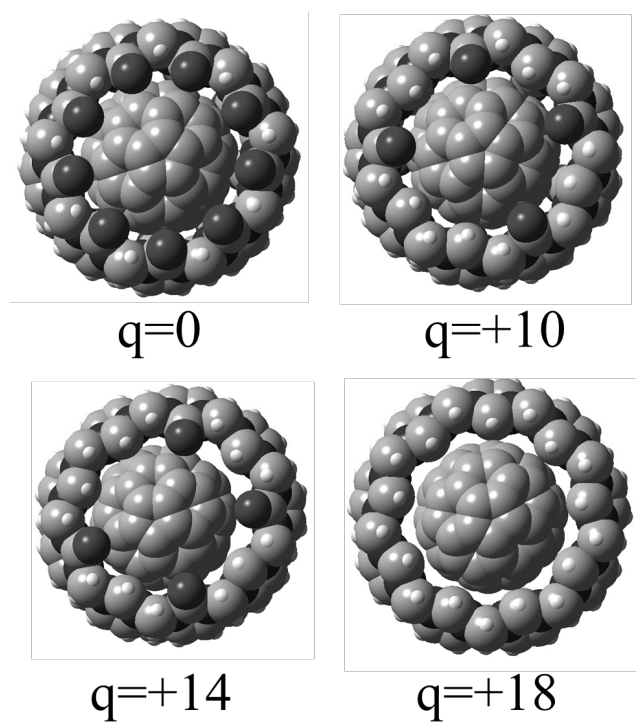

**Figure 3**. Charged and neutral host-guest complexes featuring cucurbit[9]uril and $C_{60}$ fullerene. The depicted geometries were optimized using the Berny algorithm. Oxygen atoms are black, hydrogen atoms are white, all other atoms are gray.

Figure 4 depicts a geometry optimization pathway of the $[CB[9]+C_{60}]^{14+}$ complex. During an optimization process, potential energy of the complex decreases by ca. 1000 kJ mol$^{-1}$. The energy drops down drastically during the first twenty steps. After that, it slowly approaches a final value sampling 55 additional ion-molecular configurations. This information is important, since it allows to obtain basic understanding about the potential energy surface of the host-guest complex. Stronger binding energy between non-covalently interacting species normally decreases a number of the required geometry iterations. In turn, molecular flexibility with energetically comparable conformations (such as cis- and trans-) can drastically increase this number. Depending on the purpose of optimization, the last iterations can be omitted or discarded, because energy alterations of less than one kilojoule per mole are unlikely principal for large systems and binding energy estimations. Overall, geometry optimizations of the $[CB[9]+C_{60}]^{n+}$ complexes occurred easily (for this number of degrees of freedom and realistic starting configurations), irrespective of their charge.

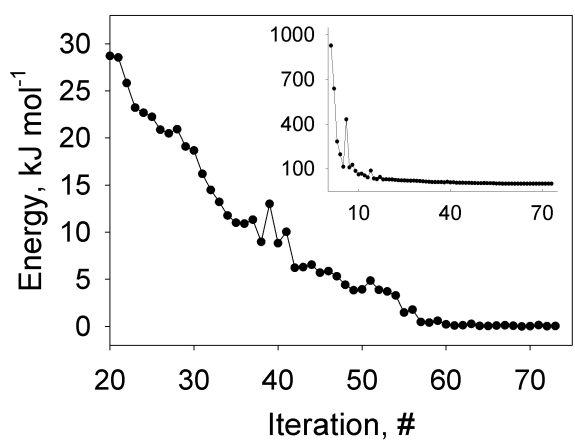

**Figure 4**. Total energy evolution during a step-by-step geometry optimization of the charged host-guest complex, $[cucurbit[9]uril+C_{60}]^{14+}$. The inset depicts total energy evolution starting from the first step. For convenience, the energy of the fully optimized geometry is assumed to be zero.

Binding energies between the $C_{60}$ fullerene and cucurbit[9]uril as a function of cationic charge of cucurbituril are summarized in Figure 5. Basis set superposition errors being ca. 10% of the binding energy were identified and excluded through the counterpoise approach. All the computed binding energies are quite significant indicating stability of the investigated host-guest complexes (Figure 3). Furthermore, a binding energy increase upon cucurbituril charge increase is clearly recorded. Fully ionized CB[9] interacts with the $C_{60}$ fullerene ca. 35% more strongly than neutral CB[9]. This difference is obviously significant. An ability to tune binding energy in this complex as a function of a positive electrostatic charge constitutes an interesting engineering opportunity. Note a very good agreement of the host-guest binding energy in the neutral complex with the result contributed before, 224 kJ mol$^{-1}$ (free energy of binding).[23] The rest of the paper is devoted to the explanation of the observed phenomenon in terms of quantum-chemical properties.

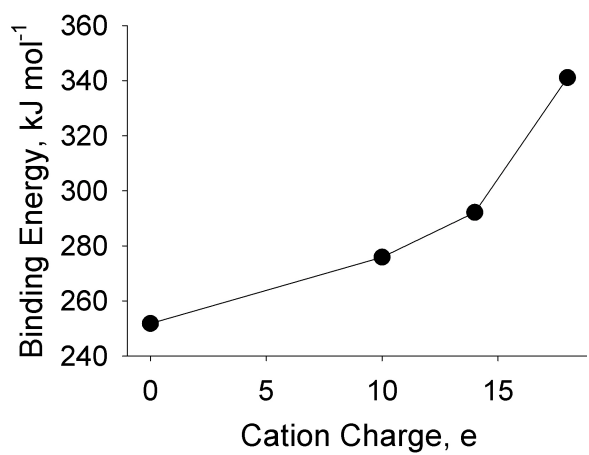

**Figure 5**. Binding energy between cucurbit[9]uril (host) and encapsulated C$_{60}$ (guest) as a function of the cucurbit[9]uril cation charge.

Although the C$_{60}$ fullerene must remain precisely neutral in the host-guest complex, Figure 6 suggests that certain deviations from neutrality do occur. These deviations clearly depend on the positive charge of the cucurbit[9]uril cation. While the total charge localized on C$_{60}$ is slightly negative in the neutral complex, it becomes positive in the cationic complexes reaching +0.27e in [CB[9]+C$_{60}$]$^{18+}$. This observation can be understood as a partial charge transfer from the C$_{60}$ fullerene to the highly charged cucurbituril ion. The charge transfer in chemistry normally indicates strengthening of mutual attraction between the involved species.

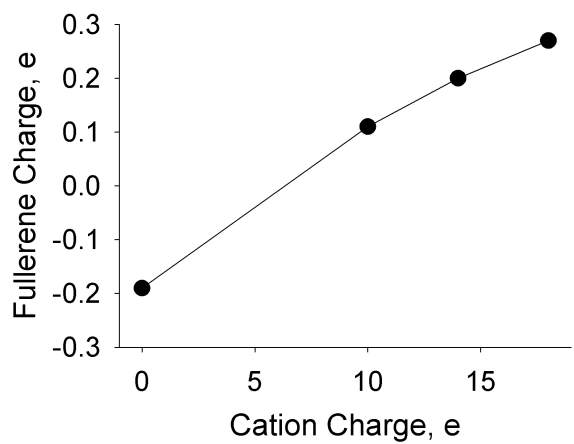

**Figure 6**. Excess or deficiency of electron density localized on the $C_{60}$ fullerene as a function of cucurbit[9]uril cation charge. We assume that $C_{60}$ remains neutral during the encapsulation process. Therefore, any deviation from zero must be regarded as an electron transfer between the host molecule and the guest molecule, which in turn increases their binding energy.

Perturbation of electron density on the $C_{60}$ fullerene in the external electric field imposed by the host molecule (CB[9]) can be efficiently characterized by the following function, $f(q) = \sqrt[2]{\sum q(C)}$, where q (C) stands for each of the sixty carbon atoms of $C_{60}$. In vacuum, an electronic structure of $C_{60}$ is unperturbed. Since carbon atoms in $C_{60}$ are chemically equivalent, their charges must ideally equal to zero and consequently $f(q) = 0$. Being submitted into an external electric field, such as one generated by neutral and ionized cucurbiturils, $C_{60}$ polarizes. Distribution of point electrostatic charges is an easiest numerical descriptor of this process providing $f(q) > 0$. Figure 7 demonstrates that $f(q)$ appears to be a clearly observable function of the cucurbituril positive charge. The higher is the charge on the cation, the larger point charges are induced in $C_{60}$. The induced charges are, in turn, responsible for an additional binding force between the host and the guest, which is electrostatic in nature. This observation is in concordance with the increased binding energy (Figure 5).

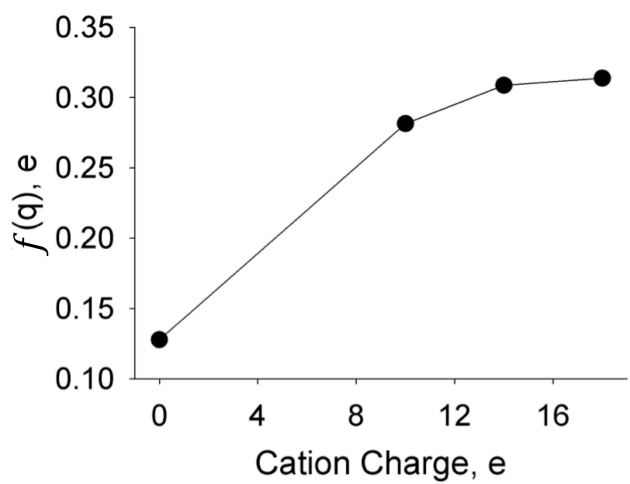

**Figure 7**. Function of point electron charges of $C_{60}$ fullerene computed as $f(q) = \sqrt[2]{\sum q(C)}$, where q (C) stands for each of the sixty carbon atoms of $C_{60}$.

Transformation of a molecule into a cation must decrease the highest occupied molecular orbital (HOMO) energy because electrons are removed from the system. We see (Figure 8) that this indeed happens. The recorded dependence is a first-order polynomial. HOMO and the lowest unoccupied molecular orbital (LUMO) are both localized on $C_{60}$. That is, the valence electronic structure of $C_{60}$ is responsible for the band gap in these systems. Cationization of cucurbit[9]uril decreases the band gap somewhat, down to ca. 1.7 eV. Note an excellent reproduction of the $C_{60}$ fullerene experimental band gap in the neutral [CB[9]+$C_{60}$] complex.[30] Sensitivity of HOMO, LUMO and their band gap in the case of $C_{60}$ fullerene indicates a strong binding in the host-guest complex.

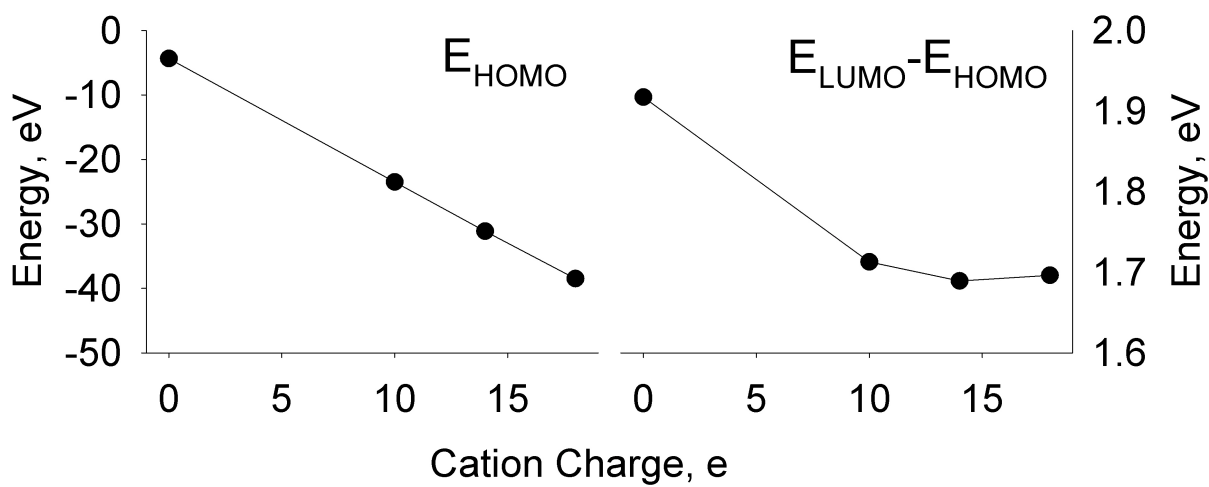

**Figure 8**. HOMO energies (left) and HOMO-LUMO band gaps (right) in the $[CB[9]+C_{60}]^{n+}$ complexes as a function of the CB[9] cation charge. The highest molecular orbitals come from $C_{60}$ being responsible for the band gap. Note an excellent reproduction of the experimental band gap for $C_{60}$ (1.9 eV) in the neutral complex.

**Conclusions**

Using molecular modeling based on hybrid density functional theory, we prove that ionized cucurbiturils bind more strongly with the $C_{60}$ fullerene. The evaluation of binding strength was performed directly via complexation energy corrected for the basis set superposition error in each system. Such an effect is possible thanks to a high electronic polarizability of fullerenes, which is, in turn, due to their strained structure. The strain increases in tiny fullerenes and decreases in giant fullerenes. For instance, one should expect very interesting effects in $C_{20}$, $C_{40}$ and their isomers. As we showed in Figure 7, $CB[n]^{n+}$ perturbs a valence electronic density of $C_{60}$ thus imposing significant partial charges on many carbon atoms. The magnitude of these charges is generally a function of n in $CB[n]^{n+}$. Consequently, the origin of cucurbituril–fullerene interaction becomes less hydrophobic and more electrostatic.

Transformation of molecules into cations and anions constitutes currently an active and promising area. See a field of task-specific room-temperature ionic liquids and ionic liquids in general for more details. Ions exhibit significantly different intermolecular interactions in relation to most compounds, as compared to their molecular precursors. Furthermore, ionic compounds mix with one another in an easier way due to the electrostatic term, which is applicable to any ion pair. This feature is largely responsible for tunability of the resulting condensed matter systems. Obviously, $CB[n]^{n+}$ cannot be used to produce a new family of ionic liquids, since these particles are too large to maintain a liquid phase ionic arrangement at ambient conditions. Nevertheless, the ions derived from cucurbiturils offer a fascinating possibility to tune stability of the host-guest complexes.

**Acknowledgments**

V.V.C. is a CAPES fellow in Brazil. T.M. and E.E.F thank Brazilian agencies FAPESP and CNPq for support.

**Authors Information**

E-mails for correspondence: vvchaban@gmail.com; fileti@gmail.com, thaciana@unifesp.br.

Tel: +55 12 3309-9573; Fax: +55 12 3921-8857.